\documentclass[%
 reprint,
 amsmath,amssymb,
 aps,
 prd,
floatfix,
]{revtex4-2}

\usepackage{graphicx} 
\usepackage{float}
\usepackage{amsmath}
\usepackage{xcolor}
\begin{document}

\title{Chiral and deconfinement properties of the QCD crossover \\
have a different volume and baryochemical potential dependence}

\author{Szabolcs Bors\'anyi$^a$, Zolt\'an Fodor $^{a,b,c,d,e}$, Jana N. Guenther$^a$, Ruben Kara$^a$, Paolo Parotto$^f$, Attila P\'asztor$^{c,g,\dagger}$, Ludovica Pirelli$^a$, Chik Him Wong $^a$}
\address{
$^a$ Department of Physics, Wuppertal University, Gaussstr.  20, D-42119, Wuppertal, Germany \\
$^b$ Pennsylvania State University, Department of Physics, State College, PA 16801, USA \\ 
$^c$ Institute  for Theoretical Physics, ELTE E\"otv\"os Lor\' and University, P\'azm\'any P. s\'et\'any 1/A, H-1117 Budapest, Hungary \\
$^d$ J\"ulich Supercomputing Centre, Forschungszentrum J\"ulich, D-52425 J\"ulich, Germany \\
$^e$ Physics Department, UCSD, San Diego, CA 92093, USA \\
$^f$ Dipartimento di Fisica, Universit\`a di Torino and INFN Torino, Via P. Giuria 1, I-10125 Torino, Italy \\
$^g$ HUN-REN-ELTE Theoretical Physics Research Group, Pázmány Péter sétány 1/A, H-1117 Budapest, Hungary \\ 
}

\date{\today}

\begin{abstract}
The crossover from hadronic to quark matter is understood to be both a
deconfinement as well as a chiral symmetry restoring transition.  Here, we
study observables related to both aspects using lattice simulations: the
Polyakov loop and its derivatives and the chiral condensate and its
derivatives.  At zero baryochemical potential, and infinite volume, the chiral
and deconfinement crossover temperatures almost agree.  However, chiral and
deconfinement related observables have a qualitatively different chemical
potential and volume dependence.  In general, deconfinement related observables
have a milder volume dependence.  Furthermore, while the deconfinement
transition appears to get broader with increasing $\mu_B$, the width as well as
the strength of the chiral transition is approximately constant.  Our results
are based on simulations at zero and imaginary chemical potentials using
4stout-improved staggered fermions with $N_\tau=12$ time-slices and physical
quark masses.
\end{abstract}

\maketitle

\section{Introduction}
The study of the phases of strongly interacting matter is an
active field of research. In particular, the phase diagram in 
the temperature $T$-baryochemical potential $\mu_B$ plane has 
garnered a lot of interest, both from the theoretical and 
experimental communities. First principle lattice QCD calculations
have shown that the transition from a hadron gas to a quark-gluon 
plasma
is a smooth crossover at zero and small chemical 
potentials~\cite{Aoki:2006we}. At larger baryochemical 
potentials, model and functional
calculations~\cite{Kovacs:2016juc,Gao:2020qsj,Fu:2021oaw,Isserstedt:2019pgx} 
predict the existence of 
a critical endpoint, where the crossover line in the 
$T-\mu_B$ plane turns into a line of first order transitions.
The position of this critical endpoint is very different 
in different model calculations. Thus, a first principle 
theory calculation, or an
experimental discovery of its location, would be a great achievement.
Indeed, the experimental discovery of this critical 
endpoint is one of the major goals of relativistic heavy ion physics~\cite{STAR:2020tga}.

\begin{figure*}[t]
\centering
\includegraphics[width=0.48\textwidth]{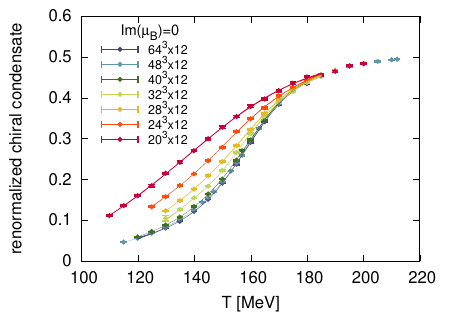}
\includegraphics[width=0.48\textwidth]{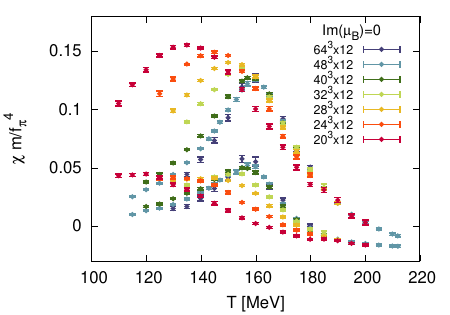}
\caption{Left panel: The renormalized chiral condensate (see eq.~\eqref{eq:chiral_cond_renorm}) as a function of temperature
for different spatial volumes. Right panel: The renormalized full and disconnected chiral susceptibility as a function of temperature
for different spatial volumes.
\label{fig:pbp_chi_mu0_vols}}
\end{figure*}

First principle lattice QCD calculation at non-zero real 
$\mu_B$ are hampered by a sign problem. Thus, most 
lattice results on 
physics at $\mu_B>0$ come from analytic continuation, 
either using lattice calculations of Taylor expansions 
around $\mu_B=0$~\cite{Gavai:2003mf,Allton:2005gk,Borsanyi:2011sw,Borsanyi:2012cr,
Bellwied:2015lba,Ding:2015fca,Bazavov:2017dus,Bazavov:2018mes,
Giordano:2019slo,Bazavov:2020bjn}, or lattice calculations at purely imaginary chemical potentials~\cite{deForcrand:2002hgr,DElia:2002tig,DElia:2009pdy,
Cea:2014xva,Bonati:2014kpa,Cea:2015cya,Bonati:2015bha,
Bellwied:2015rza,DElia:2016jqh,Gunther:2016vcp,Alba:2017mqu,
Vovchenko:2017xad,Bonati:2018nut,Borsanyi:2018grb,Bellwied:2019pxh,Borsanyi:2020fev}. While the ill-posedness of numerical
analytic continuation limits the effectiveness of
such methods, a lot has been learned by applying such methods
to the study of the phase diagram at small enough $\mu_B$.
Nevertheless, due to the ill-posed nature 
of numerical analytic continuation, such 
methods have a severe limitations. In particular,
it is exceedingly difficult to reliably 
reach large chemical
potentials with such methods.

A different set of methods is based on reweighting~\cite{
Barbour:1997ej,Fodor:2001au,Fodor:2001pe,Fodor:2004nz,deForcrand:2002pa,Alexandru:2005ix,Giordano:2020roi,Borsanyi:2021hbk, Fodor:2007vv,Endrodi:2018zda,Borsanyi:2021hbk,Borsanyi:2022soo,Borsanyi:2023tdp}. A subset of reweighting methods, sign and 
phase reweighting~\cite{Giordano:2020roi,Borsanyi:2021hbk, Borsanyi:2022soo}, have the advantage of 
giving more direct results. Thus, they have the
advantage of better reliability, when compared to
analytic continuation methods.
However, they are, at least for the
time being, restricted to smaller physical volumes, due to
the high numerical costs.
One motivation of this current work is to 
identify physical obervables with milder 
finite volume effects, which can then later
be calculated up to higher chemical potentials
with such reweighting methods.

In this work, we use the imaginary chemical potential method
to extrapolate certain physical observables to non-zero baryochemical potential $\mu_B$. We have two main goals.
First, we want to compare observables related to the 
two main theoretical aspects of the QCD transition: chiral 
symmetry, which is an $SU(2) \times SU(2)$ symmetry in 
the limit of light quarks, and confinement, which is related
to a $Z_3$ symmetry in the limit of heavy quarks. 
To study chiral dynamics, we will calculate 
the chiral condensate and chiral susceptibility. To study
confinement/deconfinement mechanics we will calculate the 
Polyakov-loop and quantities derived from it: the static
quark free energy and entropy. Second, we will look at 
the 3-volume dependence of the phase diagram. In particular,
we will quantify the volume and $\mu_B$ dependence of the 
crossover temperature, using different 
inequivalent definitions, based on both chiral symmetry
and deconfinement, and calculate parameters defining the strength and width of the
transition. 

The picture we arrive at is nuanced. At infinite volume
the crossover temperatures defined with the help of 
chiral symmetry and deconfinement are close, but not identical, and
the volume dependence of the two crossover temperatures is 
different. The definition related to deconfinement has much 
smaller finite volume effects, and the sign of the finite volume
effects is also the opposite: while the chiral crossover 
temperature increases with increasing volume, the deconfinement
temperature decreases. The width and strength 
of the transition also behave differently. Again, quantities related
to chiral symmetry show much larger finite volume effects. 
Furthermore, deconfinement related quantities show a weakening
of the crossover transition with increasing $\mu_B$, both at
small and large volumes.  This is in constrast to 
chiral observables, which show a strengthening of the transition
at smaller volumes and a roughly constant strength at large 
volumes. 

The structure of the paper is as follows. First, we discuss
the setup of our lattice simulations. Next, we discuss volume
dependence at zero chemical potential. Then, we go on to 
volume dependence at non-zero imaginary and real 
chemical potentials. We conclude with a summary and 
some speculations on the possible source of our findings.

\section{Lattice setup}
We use a tree-level Symanzik improved gauge action and four 
steps of stout smearing in the staggered fermion action, with 
$N_f=2+1+1$ flavours of dynamical quarks.
The quark masses are tuned such
that the pion and kaon masses are equal
to $135$MeV and $495$MeV. The scale is set with the 
pion decay constant $f_\pi=130.41$MeV.
Details on the action, the scale setting
and the line of constant physics can be found 
in Ref.~\cite{Bellwied:2015lba}.
We use lattices with $N_\tau=12$ timeslices. 
We use several physical volumes, with the number of spatial sites
$N_s=20,24,28,32,40,48$ and $64$ for the $\mu_B=0$ simulations. 
For three of the spatial volumes ($N_s=32,40$ and $48$) we 
also simulate at a 
purely imaginary chemical potential, with 
$\operatorname{Im} \frac{\operatorname\mu_B}{T} \frac{8}{\pi}=3,4,5,6,6.5$ and $7$. 
Throughout this work we always use a strangeness chemical
potential $\mu_S$ tuned in such a way that the expectation
value of strangeness is equal to zero. This tuning of $\mu_S$ 
is done to match conditions in heavy ion collisions, where 
the colliding nuclei
in the initial state have zero strangeness. 
Details on setting the strangeness neutrality condition 
can be found in the supplemental material of Ref.~\cite{Borsanyi:2020fev}.
In this work we only show statistical errors.

\section{Vanishing chemical potential}

\subsection{Chiral observables}

\begin{figure*}[t]
\centering
\includegraphics[width=0.48\textwidth]{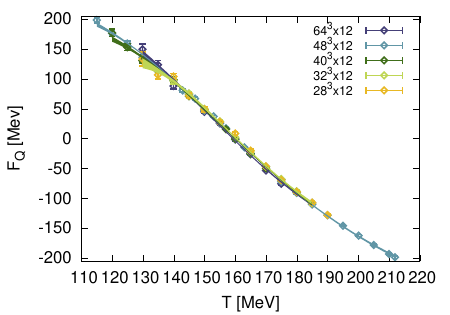}
\includegraphics[width=0.48\textwidth]{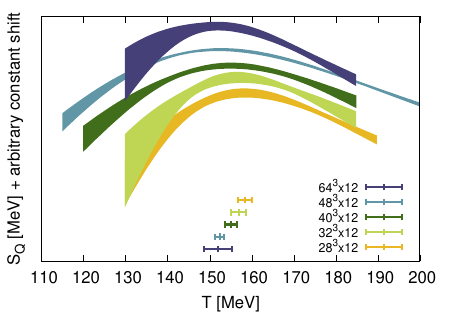}
\caption{Left panel: The static quark free energy (see eq.~\eqref{eq:FQ}) as a function of temperature
for different spatial volumes. Right panel: The static
quark entropy as a function of temperature
for different spatial volumes. In order for the curves not
to overlap, they were shifted by arbitrary amounts in the
vertical direction.
\label{fig:FQ_SQ}}
\end{figure*}

In order to study the chiral aspects of the QCD crossover
we will calculate the chiral condensate $\left\langle \bar{\psi} \psi \right\rangle$ and the chiral susceptibility $\chi$ defined respectively as:
\begin{equation}
\label{eq:chiral_cond}
\langle \bar{\psi} \psi \rangle = \frac{T}{V} \frac{\partial \log Z}{\partial m_{ud}} {\rm,} \quad\quad
\chi = \frac{T}{V} \frac{\partial^2 \log Z}{\partial m_{ud}^2} \rm.
\end{equation}
In the limit of zero quark masses $\langle \bar{\psi} \psi \rangle$ functions as the order parameter of the chiral
phase transition. In terms of the the discretized (massive) Dirac operator for the up and down quarks $M_{ud}$, the 
chiral condensate is written as:
\begin{equation}
\langle \bar{\psi} \psi \rangle = \frac{1}{2} \frac{T}{V} \left\langle \operatorname{Tr} M_{ud}^{-1}\right\rangle\rm,
\end{equation}
where $\left\langle \dots \right\rangle$ means 
taking an expectation value over
the gauge 
fields and the factor $\frac{1}{2}$ appears due to 
staggered rooting. In lattice language, 
the order parameter
is then proportional to the trace $\operatorname{Tr} M_{ud}^{-1}$.
We will also study the disconnected part of the chiral susceptibility, defined
as:
\begin{equation}
\label{eq:chiral_cond_disc}
\chi_{\rm{disc}} = \frac{1}{4} \frac{T}{V}  \left(\left\langle \left( \operatorname{Tr} M_{ud}^{-1} \right)^2\right\rangle - \left\langle \operatorname{Tr} M_{ud}^{-1} \right\rangle^2\right) \rm,
\end{equation}
where the factor of $1/4$ is present - again - due to staggered rooting. This is essentially the variance of 
$\operatorname{Tr} M_{ud}^{-1}$ in the lattice QCD language.
The
disconnected susceptibility can also be defined as a cross-derivative:
\begin{equation}
\chi_{\rm{disc}} = \frac{T}{V} \left( \frac{\partial^2}{\partial m_u \partial m_d} \log Z  \right)_{m_u=m_d} \rm,
\end{equation}
thus, it is a legitimate physical observable. We study both the full and
disconnected susceptibilities, because they may be sensitive in different ways to the distinct critical points present in the 3 dimensional QCD phase diagram spanned by the variables $T$, $\mu_B^2$ and $m_{ud}$. 

The condensate and the susceptibilities contain 
both multiplicative and additive
UV divergences. One possible definition of UV 
finite condensate and 
susceptibility is given by the following 
renormalization prescription:
\begin{equation}
\label{eq:chiral_cond_renorm}
\begin{aligned}
\langle \bar{\psi} \psi \rangle^R &= -\frac{m_{ud}}{f_\pi^4} \left[ \langle \bar{\psi} \psi \rangle_{T} - \langle \bar{\psi} \psi \rangle_{T=0}  \right] \rm, \\
\chi^R &= \frac{m_{ud}^2}{f_\pi^4} \left[  \chi_{T} - \chi_{T=0}  \right] \rm. \\
\chi_{\rm{disc}}^R &= \frac{m_{ud}^2}{f_\pi^4} \left[  \chi_{\rm{disc},T} - \chi_{\rm{disc},T=0}  \right] \rm, \\
\end{aligned}
\end{equation}
where the division by $f_\pi^4$ is there to ensure the 
quantities $\langle \bar{\psi} \psi \rangle^R, \chi^R$ and
$\chi_{\rm{disc}}^R$ are dimensionless.

The peak position of either the full or 
disconnected susceptibilities can be used to define
the chiral crossover temperature. Similarly, the maximal 
value of 
either of these susceptibilities can be used to study the
strength of the crossover transition, as they both diverge 
in the presence of true critical behavior.

The lattice results for the chiral condensate as well 
as the full and disconnected susceptibilities as functions
of the temperature are shown 
in Fig.~\ref{fig:pbp_chi_mu0_vols} for several 
lattice volumes. 

At zero temperature, chiral perturbation
theory predicts~\cite{Gerber:1988tt} the condensate to have finite volume effects of the form 
$\propto A e^{-m_\pi L} L^{-3/2}$. We have checked that this ansatz
describes our data for temperatures $T \leq 150$MeV \cite{Borsanyi:2024wuq}. The 
prefactor $A$ is related to the chiral
condensate in the chiral limit, i.e., to the topological susceptibility 
 - according to the Leutwyler-Smilga 
 relation~\cite{Leutwyler:1992yt} - which is known
to have very large cut-off effects with staggered fermions. We will thus
not present it here, from these $N_\tau=12$ lattices.

The left panel of Fig.~\ref{fig:pbp_chi_mu0_vols} shows the 
temperature dependence of the chiral condensate for different volumes, while the right panel shows the temperature dependence
of both the full and disconnected susceptibilities. 
While there are strong volume effects, both the slope of the condensate
and the maximum value of the susceptibility converges to a finite value
at large volume, consistent with a crossover.
We clearly see that the finite volume effects 
on both the condensate and the two susceptibilities decrease with increasing
temperature. This can intuitively be understood,
because the screening mass of the 
pseudoscalar channel (i.e., the smallest of the screening
masses) is increasing at temperatures 
above the crossover transition, or equivalently, 
the largest spatial correlation 
length relevant for the finite volume effects 
is decreasing. 

Near the transition temperature, both the full and disconnected 
susceptibilities show a behavior that is characteristic of a crossover transition:
their maximal value tends to a finite number
in the thermodynamic limit. Later in this manuscript, we will
study 
the chemical potential dependence of this maximal 
value, to infer whether the crossover transition 
gets stronger with increasing chemical potential, i.e., in some sense closer to 
a second order 
transition, as expected in the presence of a 
critical endpoint.

\subsection{Deconfinement observables}

We now look at properties of static test quarks,
which is a common way to characterize
confinement.

The static quark free energy~\cite{McLerran:1981pb} $F_Q$ and the static quark entropy $S_Q$ are defined as \footnote{More precisely, it is the excess
free energy from inserting a static test quark in the medium.} \cite{Bazavov:2016uvm}: 
\begin{equation}
\label{eq:FQ}
\begin{aligned}
F_Q &= - T \log \left( \frac{1}{V} \sum_{\vec{x}} |\left\langle P(\vec{x}) \right\rangle_T| \right) \\  \  & \quad + T_0 \log \left( \frac{1}{V} \sum_{\vec{x}} |\left\langle P(\vec{x}) \right\rangle_{T_0} |\right) , \\
S_Q &= - \frac{\partial F_Q}{\partial T} ,
\end{aligned}
\end{equation}
where $P(\vec{x})$ is the Polyakov loop
\begin{equation}
P(\vec{x}) = \frac{1}{3} \operatorname{Tr} \prod_{x_4=0}^{N_t-1} U_4 (\vec{x},x_4)\rm,
\end{equation}
the $U_\mu (\vec{x},x_4)$ are the usual link variables in
lattice gauge theory, $T_0$ is a reference temperature, 
and the second term in the definition of
$F_Q$ is needed to remove the additive divergence in the
free energy. The value of the 
reference temperature $T_0$, together with
Eq.~\eqref{eq:FQ} defines a 
renormalization scheme for the static quark free
energy. The subtraction has to be done at 
the same value of the lattice spacing. The 
quantity $F_Q$ is then a UV finite 
quantity, with a 
well defined continuum limit. 
Note that taking the absolute value of the (complex) Polyakov loop expectation value in Eq.~\eqref{eq:FQ} is not 
important
at vanishing or real chemical potential, as 
in both cases this expectation value is real . 
We include it in the formula, in order to keep
the same definition for arbitrary 
complex chemical potential, including a purely imaginary chemical potential, where the expectation value is no longer real and the absolute value is
needed.
The entropy $S_Q$ is estimated by 
interpolating the lattice 
results for $F_Q(T)$ and differentiating the 
interpolating function. For the interpolation 
we use a 2nd order-by-2nd order
rational function fit. 
The chi squares of all of the
fits in $T$ are acceptable. 

For pure gauge theory, where a true
first order deconfinement phase transition is present, the
static quark free energy is infinite in the confined
phase (i.e., the Polyakov loop is zero). For full QCD, 
this is no longer the case. Nevertheless, 
one can use the peak of $S_Q(T)$ (the inflection 
point of $F_Q(T)$) to define a crossover temperature.
In the limit of infinitely heavy quarks, 
the peak position of $S_Q(T)$ coincides with the 
deconfinement phase transition temperature.
The height of the peak of $S_Q(T)$ is inversely proportional
to the thermal width of the deconfinement crossover.
Our results for $F_Q(T)$ and $S_Q(T)$ are shown in
Fig.~\ref{fig:FQ_SQ} for several different 
spatial volumes.

We note that a different observable was used to define the
confinement crossover temperature in Ref.~\cite{DElia:2019iis},
where the authors introduced the susceptibility $\chi_{Q,\mu_B^2} \equiv - \left( \frac{\partial (F_Q/T)}{\partial (\mu_B/T)^2} \right)_{\mu_B=0}$, which also shows a peak at the crossover
transition. We, however, have found that we get 
statistically more precise results for $S_Q$, and therefore 
use this quantity to define the deconfinement crossover
temperature in what follows.

\begin{figure}[t]
\centering
\includegraphics[width=0.88\linewidth]{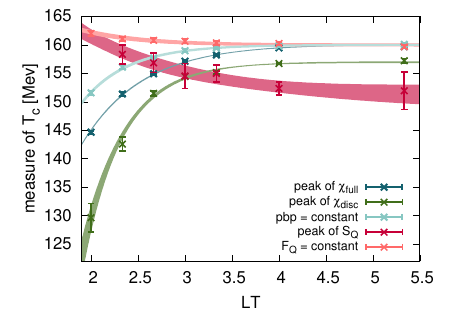}
\caption{Five different definitions of the transition temperature as functions of the simulation volume, labeled by the aspect ratio $LT=N_x/N_t$. The blue and green points refer to the peak position of the full and disconnected chiral susceptibility, respectively. In addition, we show
the temperature where the chiral condensate is constant (light blue).
In all cases, the infinite volume value is approached from below. The two other measures of $T_c$ come from the Polyakov loop, either keeping a constant value for $F_Q$ (pink), or determining the maximum of the static quark entropy $S_Q$ (purple). In both cases, the infinite volume value is approached from above. The fixed values of the chiral condensate and $F_Q$ have been set such that these match the infinite volume limit at the maximum of the full chiral susceptibility.
\label{fig:manyTc_vol}
}
\end{figure}

We are now in a position to compare the 
deconfinement and chiral symmetry related 
definitions of the crossover temperature. 
We show the different values of $T_c$ defined
as the peak of $\chi^R(T)$, the peak 
of $\chi_{disc}^R(T)$ as
well as the peak of $S_Q(T)$ in Fig.~\ref{fig:manyTc_vol}.
For comparison, we also show a curve where the chiral condensate,
or the static quark free energy is constant, with the constant 
chosen to be the infinite volume value at the crossover
temperature defined via the full chiral susceptibility.

Fig.~\ref{fig:manyTc_vol} establishes an ordering of the 
different crossover temperatures in the 
infinite volume limit:
\begin{equation}
T_c ^{(S_Q)} < T_c^{(\chi_{disc}^R)} < T_c^{\chi^R} \rm.
\end{equation}
The differences between definitions are small, amounting only to a 
few MeV.
The identification of such an ordering had never been possible in existing 
literature, due to larger error bars.
Thus, our work is an improvement
on previous work in Ref.~\cite{Bazavov:2016uvm} 
where the 
chiral and deconfinement transition temperatures were in agreement within errorbars.
Furthermore, we see a different volume dependence of the different 
crossover temperatures. Namely, the two chiral definitions lead to a $T_c$ that is
monotonically increasing with the physical volume, while the deconfinement definition
leads to a $T_c$ that is monotonically decreasing with the volume. This means that at small
enough volumes, the ordering changes, and the definition of $T_c$ based on the static
quark entropy becomes the largest.
We also note that finite volume effects on the deconfinement
based definition of $T_c$ appear to be smaller. Even on the
smallest lattice, with aspect ratio $LT=2$, the deconfinement
based definition gives a results of around $160$MeV, which is
quite close to the infinite volume value, and is much larger than the chiral symmetry based definitions.

\section{Non-vanishing chemical potential}
We now go on to discuss the phase diagram for $\mu_B>0$. First, we
briefly explain the analysis method, then we discuss the phase 
diagram, that is the volume and chemical potential dependence of
the crossover temperature, and finally, the volume and chemical
potential dependence of the strength and width
of the crossover transition. We will use the same quantities that
we used in the $\mu_B=0$ analysis.

\subsection{Observables at imaginary chemical potential}

\begin{figure}
\centering
\includegraphics[width=0.88\linewidth]{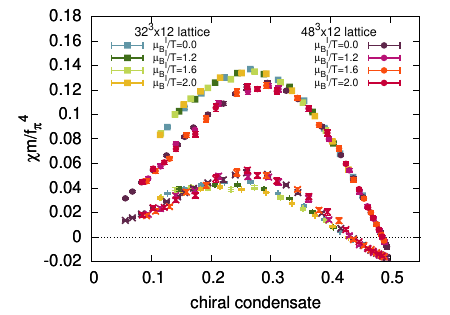}
\caption{The full (filled symbols) and disconnected (crosses) chiral susceptibilities as a function of the condensate for two different volumes and several values of the imaginary chemical potential.
\label{fig:collapse}}
\end{figure}

\begin{figure*}[t]
\centering
\includegraphics[width=0.48\textwidth]{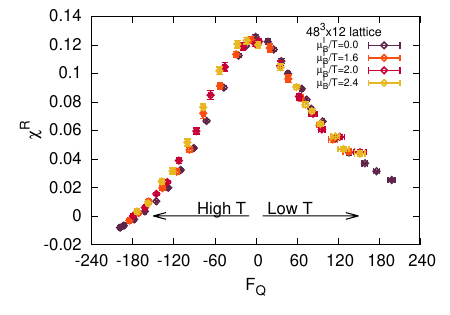}
\includegraphics[width=0.48\textwidth]{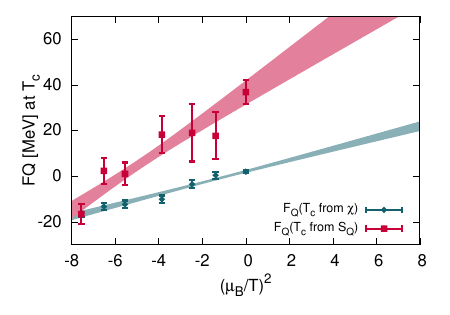}
\caption{Left: The full chiral susceptibility as a function of the static quark
free energy. Right: The value of $F_Q$ at the chiral crossover temperature (defined via the peak of the full chiral susceptibility or the static quark entropy) as a function of $\mu_B^2/T^2$. A linear extrapolation is also shown.
\label{fig:FQcollapse}}
\end{figure*}

Our extrapolation method 
to real baryochemical potential is based on 
the observation of an approximate data collapse~\cite{Borsanyi:2020fev}. The chiral susceptibility as a function of the chiral
condensate is a curve that is almost independent of the imaginary chemical potential.
In Ref.~\cite{Borsanyi:2020fev} we demonstrated this for the
full susceptibility. Here, we show it both for $\chi_{full}$ and
$\chi_{disc}$ in Fig.~\ref{fig:collapse} where both susceptibilities are shown as functions of the condensate for
several imaginary chemical potentials, for two different volumes. 
While the collapse 
curves for both susceptibilities 
are approximately independent of the imaginary chemical
potential, they significantly depend on the volume.
Also note that the susceptibilities as functions of the 
condensate are considerably simpler functions than the susceptibilities
as functions of the temperature, in the sense that they are 
well described by low order polynomials.

In light of this collapse, the extraction of 
the crossover temperature follows the 
following steps: 
\begin{enumerate}
\item determine the renormalized 
(full or disconnected) susceptibility as a function of
the condensate for several values of the imaginary baryochemical
potential;
\item find the peak position in the susceptibility 
as a function of the condensate for each value of $\operatorname{Im} \mu_B/T$ with a low order polynomial fit;
\item use an interpolation of the condensate as a function of $T$ to convert the
peak position from the condensate value to the temperature
for each $\operatorname{Im} \mu_B/T$;
\item perform a fit of $T_c(\operatorname{Im} \mu_B/T_c)$, and
use the fit to extrapolate the crossover temperature from $\mu_B^2\leq0$ to $\mu_B^2>0$.
\end{enumerate}
We note that this extrapolation method 
was also cross-checked in a calculation with truncated Dyson-Schwinger
equations, where the approximation allows for direct calculations
at $\mu_B>0$. The authors of Ref.~\cite{Bernhardt:2023ezo} found
good agreement between the direct calculations and  the extrapolated results.

If instead of the chiral condensate, one attempts to show the chiral susceptibility as a function of the static quark free energy $F_Q$, the collapse is less accurate. This is shown in 
Fig.~\ref{fig:FQcollapse}, where the left panel shows $\chi_{full}$ as a function of $F_Q$ for several imaginary chemical potentials on our $48^3 \times 12 $ lattices. To 
quantify the inaccuracy of the collapse plot we show the 
value of $F_Q$ at the crossover temperature (defined either
via $\chi_{full}$ of $S_Q$) as a function of $\mu_B^2/T^2$ 
in the right panel of the same figure, where a linear extrapolation of the value of $F_Q$ at the crossover temperature
is also shown. 
While it is not shown in the plot, to avoid clutter, we note
that the disconnected susceptibility also 
has a similar (non-perfect)
collapse feature as a function of $F_Q$. 

For the calculation of the crossover temperature defined with
$S_Q$, we simply use the same rational function fit of $F_Q(T)$
that we used for the $\mu_B=0$ case.

\begin{figure*}[t!]
\centering
\includegraphics[width=0.48\textwidth]{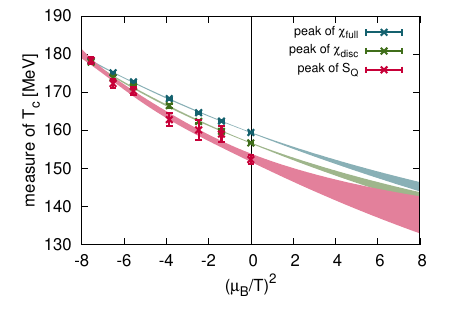}
\includegraphics[width=0.48\textwidth]{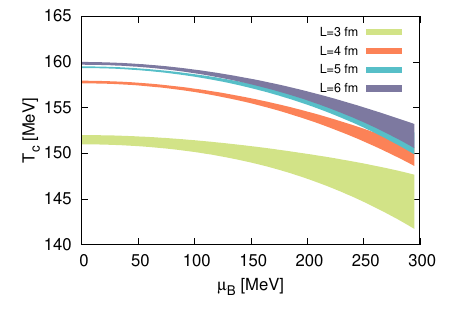}
\caption{Left panel: extrapolation of the crossover temperature
as a function of $\mu_B^2/T^2$ using the three different definitions discussed in the main text on our $48^3\times 12$ lattices. Right panel: Crossover temperature defined via the peak of $\chi_{full}$ for different lattice volumes, fixed in fermis.
\label{fig:phase_diag}}
\end{figure*}

\subsection{The phase diagram for real $\mu_B>0$}

We are now in a position to discuss the volume dependence 
of the phase diagram. We show the crossover temperature, defined
via the peak position of the full and disconnected chiral condensates, as well as the static quark entropy as a function of $\mu_B^2/T^2$ in the left panel of Fig.~\ref{fig:phase_diag} in a fixed volume,
on our $48^3 \times 12$ lattices. We see that not only $T_c(\mu_B=0)$, but also the chemical potential dependence 
is different
for the different definitions. At larger imaginary 
chemical potentials, the three definitions come closer to
each other, which might be due to
the presence of the Roberge-Weiss critical endpoint. In the
near vicinity of such a critical point, the crossover transition should get narrower, and we thus expect different
definitions of the crossover temperature to converge towards each other. This is exactly what we observe.

Here, we note that in the strangeness neutral setting employed in the present paper, the Roberge-Weiss
transition is not located at $\operatorname{Im} \mu_B/T=\pi$, but at a
slightly larger value \cite{Bonati:2014rfa}.

\begin{figure*}
\centering
\includegraphics[width=0.48\textwidth]{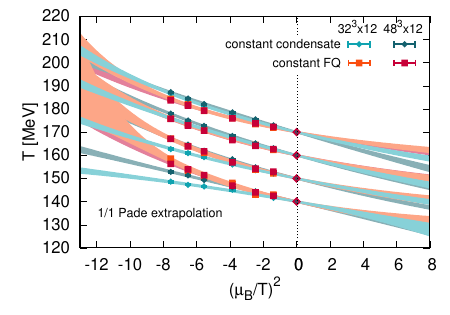}
\includegraphics[width=0.48\textwidth]{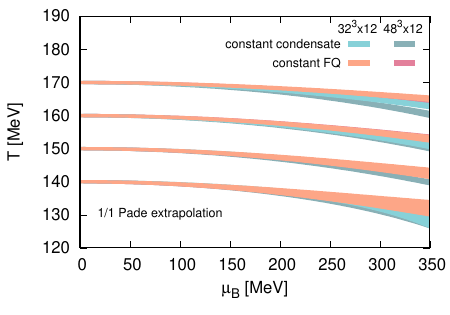}
\caption{Extrapolated contours with a constant value of the static quark free energy $F_Q$ and the renormalized chiral condensate $\left\langle \bar{\psi} \psi \right\rangle^R$ in the $T-\mu_B^2$ plane (left) and in the $T-\mu_B$ plane (right) for two different volumes.
\label{fig:const_FQ_pbp}}
\end{figure*}

We also show contours of constants 
values of the static quark free 
energy and the chiral condensate for two different
3-volumes in the $T-\mu_B^2$ plane in Fig.~\ref{fig:const_FQ_pbp}. 
It is apparent that $F_Q$ has much milder finite volume effects than
the chiral condensate, especially at imaginary chemical potentials $\mu_B^2<0$. This is somewhat surprising, since
at imaginary chemical potentials, one might expect $F_Q$ to have
strong finite-volume effects, being closely related to the Polyakov loop, which is the order parameter for the Roberge-Weiss transition at imaginary chemical potentials.

\subsection{The strength of the crossover for real $\mu_B \neq 0$}

\begin{figure}
\centering
\includegraphics[width=0.88\linewidth]{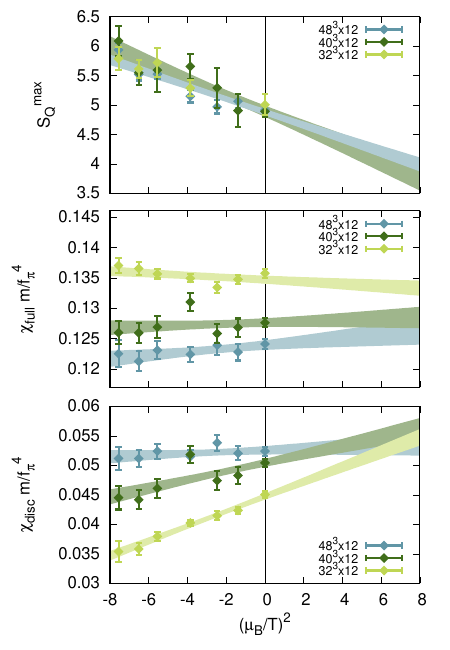}
\caption{Extrapolation of the maximal value of 
the full and disconnected chiral
susceptibilities as a function of $\mu_B^2$ 
for three different volumes.
\label{fig:strength}}
\end{figure}

A very interesting question, with more phenomenological implications, is whether the crossover line
turns into a line of first order transitions at a critical endpoint at
some real value of the baryochemical potential $\mu_B$.
If it does, it is expected that the crossover transition becomes
narrower and stronger at larger $\mu_B$, at least in the
vicinity of the critical endpoint. Here, we discuss some measures
of the width or strength of the transition as a function
of $\mu_B$ for small chemical potentials. 

In Fig.~\ref{fig:strength} we show the values 
of $S_Q$, $\chi^R$ and $\chi^R_{disc}$ at the 
crossover temperature as functions of $\mu_B^2/T^2$, for 
two different physical 3-volumes. 
The points with error bars are our direct lattice
results at zero and purely imaginary chemical potentials,
while the bands show linear extrapolations in $\mu_B^2/T^2$.
All three quantities should diverge at the critical endpoint in the infinite volume limit. 
Thus, with increasing volume, one expects these quantities 
to grow in the vicinity of the critical endpoint.
The only one of these quantities showing a rise at larger $\mu_B^2$ is $\chi^R_{disc}$, while 
$\chi^R$ remains approximately constant up to our largest volume, 
and $S_Q$ decreases with increasing $\mu_B^2$. From the behavior of $\chi^R_{disc}$ alone, one might be tempted to conclude
that the crossover transition gets stronger 
at larger $\mu_B$, which would be a signal of the coveted
critical endpoint. However, the lack of a similar behavior in the other two quantities - $S_Q$ and $\chi^R$ makes the interpretation of the physical picture uncertain.

Furthermore, we once again see that the deconfinement related quantity ($S_Q$) has
milder finite volume effects than the quantities related to 
chiral symmetry restoration ($\chi^R$ and $\chi^R_{disc}$).
Just like in the case of $F_Q$ above, the small volume dependence of $S_Q$
at imaginary chemical potentials is surprising, since one
expects large finite volume effects due to the Roberge-Weiss 
endpoint. Apparently the imaginary chemical 
potentials considered in our study are not 
close enough the the Roberge-Weiss critical
endpoint for this to be the case.

The disconnected susceptibility shows the most 
interesting behavior: an increasing trend with 
increasing $\mu_B^2$, especially for the two lower volumes. If the coveted 
QCD critical endpoint exists, such a behavior could be due to the
critical region shrinking with increasing physical volume. 
(I.e., a smaller volume is more tolerant to the mistuning of the parameters of 
a system, and criticality can be observed from farther away.)
Due to the lack of such a signal in $\chi^R$ and $S_Q$, we
would be cautious about making such an interpretation, but the behavior of
$\chi^R_{disc}$ certainly is suggestive and 
warrants further investigation.

\section{Discussion}

In this work we investigated the volume and chemical potential dependence of the location and the strength of the QCD transition, by studying observables related to either deconfinement or chiral symmetry restoration.

At vanishing chemical potential, we found - in the thermodynamic limit - the following ordering of the crossover temperatures: $T_c^{(S_Q)} < T_c^{(\chi_{disc}^R)} < T_c^{(\chi^R)}$. Hence, we found the deconfinement crossover temperatures to be slightly below the chiral crossover temperature. 
Furthermore, we found that the deconfinement 
crossover temperature and the chiral crossover temperatures
have a different sign for their volume dependence. While with increasing volume $T_c^{(S_Q)}$
decreases, $T_c^{(\chi_{disc}^R)}$ and $T_c^{(\chi^R)}$ both increase. 

Curves of constant heavy quark free energy (related to confinement) and constant chiral condensate (related to chiral symmetry breaking) in the $T-\mu_B^2/T^2$ plane are similar, 
but not identical. In particular, the curvature of these lines
is about the same at $T_c(\mu_B=0)$, but not below or above. 

In general, we find that physical quantities related to confinement dynamics
appear to have much milder finite volume effects, when compared
to quantities related to chiral symmetry. This is true at both zero and non-zero baryochemical potential, be it imaginary (simulated) or real (extrapolated).
Thus, confinement-deconfinement mechanics will be easier to study in settings where larger volumes are too expensive, which
is the case, e.g., with reweighting techniques.

Finally, we found that the volume and $\mu_B$ dependence of the maximal value of $\chi^R$, $\chi^R_{disc}$ and $S_Q$ are non-trivial, and a clear physical picture cannot be drawn at the moment. However, the disconnected susceptibility $\chi^R_{disc}$ in particular shows suggestive behavior, which is promising in light of
future theoretical searches for the QCD critical point.

\section*{Acknowledgements}
The project was supported by the BMBF Grant
No. 05P21PXFCA. This work is also supported by the
MKW NRW under the funding code NW21-024-A. Further
funding was received from the DFG under the Project
No. 496127839. This work was also supported by the
Hungarian National Research, Development and Innovation
Office, NKFIH Grant No. KKP126769.
This work was also supported by the NKFIH excellence
grant TKP2021{\textunderscore}NKTA{\textunderscore}64.
The project was also supported by the Hungarian National Research,
Development and Innovation
Office under Project No. FK 147164.
The authors gratefully acknowledge the Gauss Centre for
Supercomputing e.V. (\url{www.gauss-centre.eu}) for funding
this project by providing computing time on the GCS
Supercomputers Juwels-Booster at Juelich Supercomputer
Centre and HAWK at H\"ochstleistungsrechenzentrum Stuttgart.

\bibliographystyle{unsrt} 
\bibliography{thermo}

\begin{thebibliography}{10}

\bibitem{Aoki:2006we}
Y.~Aoki, G.~Endrodi, Z.~Fodor, S.D. Katz, and K.K. Szabo.
\newblock {The Order of the quantum chromodynamics transition predicted by the
  standard model of particle physics}.
\newblock {\em Nature}, 443:675--678, 2006.

\bibitem{Kovacs:2016juc}
Peter Kov\'acs, Zsolt Sz\'ep, and Gy\"orgy Wolf.
\newblock {Existence of the critical endpoint in the vector meson extended
  linear sigma model}.
\newblock {\em Phys. Rev. D}, 93(11):114014, 2016.

\bibitem{Gao:2020qsj}
Fei Gao and Jan~M. Pawlowski.
\newblock {QCD phase structure from functional methods}.
\newblock {\em Phys. Rev. D}, 102(3):034027, 2020.

\bibitem{Fu:2021oaw}
Wei-jie Fu, Xiaofeng Luo, Jan~M. Pawlowski, Fabian Rennecke, Rui Wen, and Shi
  Yin.
\newblock {Hyper-order baryon number fluctuations at finite temperature and
  density}.
\newblock {\em Phys. Rev. D}, 104(9):094047, 2021.

\bibitem{Isserstedt:2019pgx}
Philipp Isserstedt, Michael Buballa, Christian~S. Fischer, and Pascal~J.
  Gunkel.
\newblock {Baryon number fluctuations in the QCD phase diagram from
  Dyson-Schwinger equations}.
\newblock {\em Phys. Rev.}, D100(7):074011, 2019.

\bibitem{STAR:2020tga}
J.~Adam et~al.
\newblock {Nonmonotonic Energy Dependence of Net-Proton Number Fluctuations}.
\newblock {\em Phys. Rev. Lett.}, 126(9):092301, 2021.

\bibitem{Gavai:2003mf}
Rajiv~V. Gavai and Sourendu Gupta.
\newblock {Pressure and nonlinear susceptibilities in QCD at finite chemical
  potentials}.
\newblock {\em Phys.Rev.}, D68:034506, 2003.

\bibitem{Allton:2005gk}
C.R. Allton, M.~Doring, S.~Ejiri, S.J. Hands, O.~Kaczmarek, et~al.
\newblock {Thermodynamics of two flavor QCD to sixth order in quark chemical
  potential}.
\newblock {\em Phys.Rev.}, D71:054508, 2005.

\bibitem{Borsanyi:2011sw}
Szabolcs Borsanyi, Zoltan Fodor, Sandor~D. Katz, Stefan Krieg, Claudia Ratti,
  et~al.
\newblock {Fluctuations of conserved charges at finite temperature from lattice
  QCD}.
\newblock {\em JHEP}, 1201:138, 2012.

\bibitem{Borsanyi:2012cr}
Sz. Borsanyi, G.~Endrodi, Z.~Fodor, S.D. Katz, S.~Krieg, et~al.
\newblock {QCD equation of state at nonzero chemical potential: continuum
  results with physical quark masses at order $mu^2$}.
\newblock {\em JHEP}, 1208:053, 2012.

\bibitem{Bellwied:2015lba}
R.~Bellwied, S.~Borsanyi, Z.~Fodor, S.~D. Katz, A.~Pasztor, C.~Ratti, and K.~K.
  Szabo.
\newblock {Fluctuations and correlations in high temperature QCD}.
\newblock {\em Phys. Rev.}, D92(11):114505, 2015.

\bibitem{Ding:2015fca}
H.~T. Ding, Swagato Mukherjee, H.~Ohno, P.~Petreczky, and H.~P. Schadler.
\newblock {Diagonal and off-diagonal quark number susceptibilities at high
  temperatures}.
\newblock {\em Phys. Rev.}, D92(7):074043, 2015.

\bibitem{Bazavov:2017dus}
A.~Bazavov et~al.
\newblock {The QCD Equation of State to $\mathcal{O}(\mu_B^6)$ from Lattice
  QCD}.
\newblock {\em Phys. Rev.}, D95(5):054504, 2017.

\bibitem{Bazavov:2018mes}
A.~Bazavov et~al.
\newblock {Chiral crossover in QCD at zero and non-zero chemical potentials}.
\newblock {\em Physics Letters B}, 795:15--21, Aug 2019.

\bibitem{Giordano:2019slo}
Matteo Giordano and Attila Psztor.
\newblock {Reliable estimation of the radius of convergence in finite density
  QCD}.
\newblock {\em Phys. Rev.}, D99(11):114510, 2019.

\bibitem{Bazavov:2020bjn}
A.~Bazavov et~al.
\newblock {Skewness, kurtosis and the 5th and 6th order cumulants of net
  baryon-number distributions from lattice QCD confront high-statistics STAR
  data}.
\newblock {\em Physical Review D}, 101(7), Apr 2020.

\bibitem{deForcrand:2002hgr}
Philippe de~Forcrand and Owe Philipsen.
\newblock {The QCD phase diagram for small densities from imaginary chemical
  potential}.
\newblock {\em Nucl. Phys.}, B642:290--306, 2002.

\bibitem{DElia:2002tig}
Massimo D'Elia and Maria-Paola Lombardo.
\newblock {Finite density QCD via imaginary chemical potential}.
\newblock {\em Phys. Rev.}, D67:014505, 2003.

\bibitem{DElia:2009pdy}
Massimo D'Elia and Francesco Sanfilippo.
\newblock {Thermodynamics of two flavor QCD from imaginary chemical
  potentials}.
\newblock {\em Phys. Rev.}, D80:014502, 2009.

\bibitem{Cea:2014xva}
Paolo Cea, Leonardo Cosmai, and Alessandro Papa.
\newblock {Critical line of 2+1 flavor QCD}.
\newblock {\em Phys. Rev.}, D89(7):074512, 2014.

\bibitem{Bonati:2014kpa}
Claudio Bonati, Philippe de~Forcrand, Massimo D'Elia, Owe Philipsen, and
  Francesco Sanfilippo.
\newblock {The chiral phase transition in two-flavor QCD from imaginary
  chemical potential}.
\newblock {\em Physical Review D}, 90(7), Oct 2014.

\bibitem{Cea:2015cya}
Paolo Cea, Leonardo Cosmai, and Alessandro Papa.
\newblock {Critical line of 2+1 flavor QCD: Toward the continuum limit}.
\newblock {\em Phys. Rev.}, D93(1):014507, 2016.

\bibitem{Bonati:2015bha}
Claudio Bonati, Massimo D'Elia, Marco Mariti, Michele Mesiti, Francesco Negro,
  and Francesco Sanfilippo.
\newblock {Curvature of the chiral pseudocritical line in QCD: Continuum
  extrapolated results}.
\newblock {\em Phys. Rev.}, D92(5):054503, 2015.

\bibitem{Bellwied:2015rza}
R.~Bellwied, S.~Borsanyi, Z.~Fodor, J.~G{\"u}nther, S.~D. Katz, C.~Ratti, and
  K.~K. Szabo.
\newblock {The QCD phase diagram from analytic continuation}.
\newblock {\em Phys. Lett.}, B751:559--564, 2015.

\bibitem{DElia:2016jqh}
Massimo D'Elia, Giuseppe Gagliardi, and Francesco Sanfilippo.
\newblock {Higher order quark number fluctuations via imaginary chemical
  potentials in $N_f=2+1$ QCD}.
\newblock {\em Phys. Rev.}, D95(9):094503, 2017.

\bibitem{Gunther:2016vcp}
J.~Gunther, R.~Bellwied, S.~Borsanyi, Z.~Fodor, S.~D. Katz, A.~Pasztor, and
  C.~Ratti.
\newblock {The QCD equation of state at finite density from analytical
  continuation}.
\newblock {\em EPJ Web Conf.}, 137:07008, 2017.

\bibitem{Alba:2017mqu}
Paolo Alba et~al.
\newblock {Constraining the hadronic spectrum through QCD thermodynamics on the
  lattice}.
\newblock {\em Phys. Rev.}, D96(3):034517, 2017.

\bibitem{Vovchenko:2017xad}
Volodymyr Vovchenko, Attila Pasztor, Zoltan Fodor, Sandor~D. Katz, and Horst
  Stoecker.
\newblock {Repulsive baryonic interactions and lattice QCD observables at
  imaginary chemical potential}.
\newblock {\em Phys. Lett.}, B775:71--78, 2017.

\bibitem{Bonati:2018nut}
Claudio Bonati, Massimo D'Elia, Francesco Negro, Francesco Sanfilippo, and
  Kevin Zambello.
\newblock {Curvature of the pseudocritical line in QCD: Taylor expansion
  matches analytic continuation}.
\newblock {\em Phys. Rev.}, D98(5):054510, 2018.

\bibitem{Borsanyi:2018grb}
Szabolcs Borsanyi, Zoltan Fodor, Jana~N. Guenther, Sandor~K. Katz, Kalman~K.
  Szabo, Attila Pasztor, Israel Portillo, and Claudia Ratti.
\newblock {Higher order fluctuations and correlations of conserved charges from
  lattice QCD}.
\newblock {\em JHEP}, 10:205, 2018.

\bibitem{Bellwied:2019pxh}
Rene Bellwied, Szabolcs Borsanyi, Zoltan Fodor, Jana~N. Guenther, Jacquelyn
  Noronha-Hostler, Paolo Parotto, Attila Pasztor, Claudia Ratti, and Jamie~M.
  Stafford.
\newblock {Off-diagonal correlators of conserved charges from lattice QCD and
  experiment}.
\newblock {\em Physical Review D}, 101(3), Feb 2020.

\bibitem{Borsanyi:2020fev}
Szabolcs Borsanyi, Zoltan Fodor, Jana~N. Guenther, Ruben Kara, Sandor~D. Katz,
  Paolo Parotto, Attila Pasztor, Claudia Ratti, and Kalman~K. Szabo.
\newblock {The QCD crossover at finite chemical potential from lattice
  simulations}.
\newblock {\em Phys. Rev. Lett.}, 125:052001, 2020.

\bibitem{Barbour:1997ej}
Ian~M. Barbour, Susan~E. Morrison, Elyakum~G. Klepfish, John~B. Kogut, and
  Maria-Paola Lombardo.
\newblock {Results on finite density QCD}.
\newblock {\em Nucl. Phys. Proc. Suppl.}, 60A:220--234, 1998.
\newblock [,220(1997)].

\bibitem{Fodor:2001au}
Z.~Fodor and S.D. Katz.
\newblock {A New method to study lattice QCD at finite temperature and chemical
  potential}.
\newblock {\em Phys.Lett.}, B534:87--92, 2002.

\bibitem{Fodor:2001pe}
Z.~Fodor and S.D. Katz.
\newblock {Lattice determination of the critical point of QCD at finite T and
  mu}.
\newblock {\em JHEP}, 0203:014, 2002.

\bibitem{Fodor:2004nz}
Z.~Fodor and S.D. Katz.
\newblock {Critical point of QCD at finite T and mu, lattice results for
  physical quark masses}.
\newblock {\em JHEP}, 0404:050, 2004.

\bibitem{deForcrand:2002pa}
P.~de~Forcrand, S.~Kim, and T.~Takaishi.
\newblock {QCD simulations at small chemical potential}.
\newblock {\em Nucl. Phys. B Proc. Suppl.}, 119:541--543, 2003.

\bibitem{Alexandru:2005ix}
Andrei Alexandru, Manfried Faber, Ivan Horvath, and Keh-Fei Liu.
\newblock {Lattice QCD at finite density via a new canonical approach}.
\newblock {\em Phys. Rev.}, D72:114513, 2005.

\bibitem{Giordano:2020roi}
Matteo Giordano, Kornel Kapas, Sandor~D. Katz, Daniel Nogradi, and Attila
  Pasztor.
\newblock {New approach to lattice QCD at finite density; results for the
  critical end point on coarse lattices}.
\newblock {\em JHEP}, 05:088, 2020.

\bibitem{Borsanyi:2021hbk}
Szabolcs Borsanyi, Zoltan Fodor, Matteo Giordano, Sandor~D. Katz, Daniel
  Nogradi, Attila Pasztor, and Chik~Him Wong.
\newblock {Lattice simulations of the QCD chiral transition at real baryon
  density}.
\newblock {\em Phys. Rev. D}, 105(5):L051506, 2022.

\bibitem{Fodor:2007vv}
Zoltan Fodor, Sandor~D. Katz, and Christian Schmidt.
\newblock {The Density of states method at non-zero chemical potential}.
\newblock {\em JHEP}, 0703:121, 2007.

\bibitem{Endrodi:2018zda}
G.~Endrodi, Z.~Fodor, S.~D. Katz, D.~Sexty, K.~K. Szabo, and Cs. Torok.
\newblock {Applying constrained simulations for low temperature lattice QCD at
  finite baryon chemical potential}.
\newblock {\em Phys. Rev.}, D98(7):074508, 2018.

\bibitem{Borsanyi:2022soo}
Szabolcs Borsanyi, Zoltan Fodor, Matteo Giordano, Jana~N. Guenther, Sandor~D.
  Katz, Attila Pasztor, and Chik~Him Wong.
\newblock {Equation of state of a hot-and-dense quark gluon plasma: Lattice
  simulations at real \ensuremath{\mu}B vs extrapolations}.
\newblock {\em Phys. Rev. D}, 107(9):L091503, 2023.

\bibitem{Borsanyi:2023tdp}
Szabolcs Borsanyi, Zoltan Fodor, Matteo Giordano, Jana~N. Guenther, Sandor~D.
  Katz, Attila Pasztor, and Chik~Him Wong.
\newblock {Can rooted staggered fermions describe nonzero baryon density at low
  temperatures?}
\newblock 8 2023.

\bibitem{Gerber:1988tt}
P.~Gerber and H.~Leutwyler.
\newblock {Hadrons Below the Chiral Phase Transition}.
\newblock {\em Nucl.Phys.}, B321:387, 1989.

\bibitem{Borsanyi:2024wuq}
Szabolcs Bors\'anyi, Zolt\'an Fodor, Jana~N. Guenther, Ruben Kara, Paolo
  Parotto, Attila P\'asztor, and Chik~Him Wong.
\newblock {Finite volume effects near the chiral crossover}.
\newblock 1 2024.

\bibitem{Leutwyler:1992yt}
H.~Leutwyler and Andrei~V. Smilga.
\newblock {Spectrum of Dirac operator and role of winding number in QCD}.
\newblock {\em Phys. Rev.}, D46:5607--5632, 1992.

\bibitem{McLerran:1981pb}
Larry~D. McLerran and Benjamin Svetitsky.
\newblock {Quark Liberation at High Temperature: A Monte Carlo Study of SU(2)
  Gauge Theory}.
\newblock {\em Phys. Rev. D}, 24:450, 1981.

\bibitem{Note1}
More precisely, it is the excess free energy from inserting a static test quark
  in the medium.

\bibitem{Bazavov:2016uvm}
A.~Bazavov, N.~Brambilla, H.~T. Ding, P.~Petreczky, H.~P. Schadler, A.~Vairo,
  and J.~H. Weber.
\newblock {Polyakov loop in 2+1 flavor QCD from low to high temperatures}.
\newblock {\em Phys. Rev.}, D93(11):114502, 2016.

\bibitem{DElia:2019iis}
Massimo D'Elia, Francesco Negro, Andrea Rucci, and Francesco Sanfilippo.
\newblock {Dependence of the static quark free energy on $\mu_B$ and the
  crossover temperature of $N_f = 2+1$ QCD}.
\newblock {\em Phys. Rev. D}, 100(5):054504, 2019.

\bibitem{Bernhardt:2023ezo}
Julian Bernhardt and Christian~S. Fischer.
\newblock {From imaginary to real chemical potential QCD with functional
  methods}.
\newblock {\em Eur. Phys. J. A}, 59(8):181, 2023.

\bibitem{Bonati:2014rfa}
Claudio Bonati, Massimo D'Elia, Marco Mariti, Michele Mesiti, Francesco Negro,
  et~al.
\newblock {Curvature of the chiral pseudocritical line in QCD}.
\newblock {\em Phys.Rev.}, D90(11):114025, 2014.

\end{thebibliography}

\end{document}